\documentclass[twoside,english,journal, twocolumn]{IEEEtran}
\usepackage[T1]{fontenc}
\usepackage[latin9]{inputenc}
\usepackage{amsmath}
\usepackage{graphicx}
\usepackage{setspace}
\usepackage{amssymb}

\providecommand{\tabularnewline}{\\}

\usepackage{color, graphicx}
\usepackage{cite}
\usepackage{epsfig, psfrag}
\usepackage{subfigure}
\newtheorem{theorem}{Theorem}[section]

\usepackage{babel}

\begin{document}

\title{Communication in a Poisson Field of Interferers -- Part~I: Interference
Distribution and Error Probability}

\author{Pedro C.\ Pinto, \IEEEmembership{Student Member, IEEE,} and Moe
Z.\ Win, \IEEEmembership{Fellow, IEEE}%
\thanks{Manuscript received July~3, 2006; revised Apr~29, 2007; accepted
Mar~20, 2008. The editor coordinating the review of this paper and
approving it for publication is D.~Dardari. This research was supported,
in part, by the Portuguese Science and Technology Foundation under
grant SFRH-BD-17388-2004, the Charles Stark Draper Laboratory Robust
Distributed Sensor Networks Program, the Office of Naval Research
Young Investigator Award N00014-03-1-0489, and the National Science
Foundation under Grant ANI-0335256. This paper was presented, in part,
at the IEEE Conference on Information Sciences and Systems, Princeton,
NJ, March 2006.%
}%
\thanks{P.~C.~Pinto and M.~Z.~Win are with the Laboratory for Information
and Decision Systems (LIDS), Massachusetts Institute of Technology,
Room~\mbox{32-D674}, 77~Massachusetts Avenue, Cambridge, MA 02139,
USA (\mbox{e-mail}: \texttt{ppinto@mit.edu}, \texttt{moewin@mit.edu}).%
}%
\thanks{Digital Object Identifier 10.1109/TWC.2008.XXXXXXX%
}}

\maketitle

\thispagestyle{empty}

\newpage

\setcounter{page}{1}
\begin{abstract}
\noindent We present a mathematical model for communication subject
to both network interference and noise. We introduce a framework where
the interferers are scattered according to a spatial Poisson process,
and are operating asynchronously in a wireless environment subject
to path loss, shadowing, and multipath fading. We consider both cases
of slow and fast-varying interferer positions. The paper is comprised
of two separate parts. In Part~I, we determine the distribution of
the aggregate network interference at the output of a linear receiver.
We characterize the error performance of the link, in terms of average
and outage probabilities. The proposed model is valid for any linear
modulation scheme (e.g.,~$\mbox{$M$-ary}$ phase shift keying or
$\mbox{$M$-ary}$ quadrature amplitude modulation), and captures all
the essential physical parameters that affect network interference.
Our work generalizes the conventional analysis of communication in
the presence of additive white Gaussian noise and fast fading, allowing
the traditional results to be extended to include the effect of network
interference. In Part~II of the paper, we derive the capacity of
the link when subject to network interference and noise, and characterize
the spectrum of the aggregate interference.\end{abstract}

\section{Introduction\label{sec:Introduction}}

\PARstart{I}{n} a wireless network composed of many spatially scattered
nodes, there are several fundamental impairments that constrain the
communication between nodes, including \emph{thermal noise} and \emph{network
interference}. Thermal noise is introduced by the receiver electronics
and is usually modeled as additive white Gaussian noise (AWGN), which
constitutes a good approximation in most cases. Interference, on the
other hand, is due to signals radiated by other transmitters, which
undesirably affect receiver nodes in the same or in a different network.
For simplicity, interference is typically approximated by AWGN with
some given power~\cite{VitJac:75}. However, this elementary model
does not completely capture the physical parameters that affect interference,
namely: 1)~the spatial distribution of nodes scattered in the network;
2)~the transmission characteristics of nodes, such as modulation,
power, and synchronization; and 3)~the propagation characteristics
of the medium, such as path loss, shadowing, and multipath fading.
If, instead, a spatial Poisson process is used to model the user positions,
then all these parameters are easily accounted for, and appear explicitly
in the resulting performance expressions. 

The application of the Poisson field model to cellular networks was
investigated in \cite{Sou:92} and later advanced in \cite{IloHatVen:98}.
However, these papers either ignore random propagation effects (such
as shadowing and multipath fading), or restrict the analysis of error
probability in non-coherent FSK modulations. In other related work~\cite{YanPet:03},
it is assumed that the different interferers are synchronized at the
symbol or slot level, which may be unrealistic in most situations.
In \cite{SalZan:07}, the authors choose a different approach and
restrict the node locations to a disk or ring in the two-dimensional
plane. Although this ensures that the number of interferers is finite,
it complicates the analysis and does not provide useful insights into
the effects of network interference. In \cite{GlaGre:83,GioChiWin:J05,GioChiDar:06},
the authors analyze coexistence issues in wireless networks, but consider
only a small, fixed number of interferers. Lastly, none of the mentioned
studies attempts a system characterization that incorporates various
metrics such as error probability, channel capacity, and power spectral
density.

In this two-part paper, we introduce a more realistic framework where
the interferers are scattered according to a spatial Poisson process,
and are operating asynchronously in a wireless environment subject
to path loss, shadowing, and multipath fading~\cite{PinWin:C06,PinWin:J07,PinChoGioChiWin:C06,WinPinGioChiShe:C06}.
We specifically address the cases of slow and fast-varying interferer
positions. In Part~I of the paper, we determine the statistical distribution
of the aggregate network interference at the output of a linear receiver,
located anywhere in the two-dimensional plane. We provide expressions
for the error performance of the link (in terms of average and outage
probabilities), which are valid for any linear modulation scheme.
We then quantify these metrics as a function of various important
system parameters, such as the signal-to-noise ratio (SNR), interference-to-noise
ratio (INR), path loss exponent, and spatial density of the interferers.
Our analysis clearly shows how the system performance depends on these
parameters, thereby providing insights that may be of value to the
network designer. In Part~II of the paper~\cite{PinWin:J09a}, we
derive the capacity of the link when subject to network interference
and noise, and characterize the spectrum of the aggregate interference.

This paper is organized as follows. Section~\ref{sec:System-model}
describes the system model. Section~\ref{sec:Interference-Distribution}
derives the representation and distribution of the aggregate interference.
Section~\ref{sec:Error-probability} analyzes the error performance
of the system, and gives plots to illustrate its dependence on important
network parameters. Section~\ref{sec:Summary} concludes the paper
and summarizes important findings.

\section{System Model\label{sec:System-model}}

\subsection{Spatial Distribution of the Nodes}

We model the spatial distribution of the nodes according to a homogeneous
Poisson point process in the two-dimensional infinite plane. Typically,
the terminal positions are unknown to the network designer a priori,
so we may as well treat them as completely random according to a spatial
Poisson process.%
\footnote{The spatial Poisson process is a natural choice in such situation
because, given that a node is inside a region~$\mathcal{R}$, the
probability density function (p.d.f.) of its position is conditionally
uniform over $\mathcal{R}$.%
} Then, the probability~$\mbox{$\mathbb{P}\{n$ in $\mathcal{R}\}$}$
of $n$~nodes being inside a region~$\mathcal{R}$ (not necessarily
connected) depends only on the total area~$\mathcal{A}$ of the region,
and is given by~\cite{Kin:93}\[
\mathbb{P}\{n\textrm{ in }\mathcal{R}\}=\frac{(\lambda\mathcal{A})^{n}}{n!}e^{-\lambda\mathcal{A}},\quad n\geq0,\]
where $\lambda$ is the (constant) spatial density of interfering
nodes, in nodes per unit area. We define the \emph{interfering nodes}
to be the set of terminals which are transmitting within the frequency
band of interest, during the time interval of interest, and hence
are effectively contributing to the total interference. Then, irrespective
of the network topology (e.g.,~point-to-point or broadcast) or multiple-access
technique (e.g.,~time or frequency hopping), the above model depends
only on the density~$\lambda$ of interfering nodes.%
\footnote{Time and frequency hopping can be easily accommodated in this model,
using the splitting property of Poisson processes~\cite{BerTsi:02}
to obtain the \emph{effective} density of nodes that contribute to
the interference.%
}

\begin{figure}
\begin{centering}
\scalebox{0.6}{\input{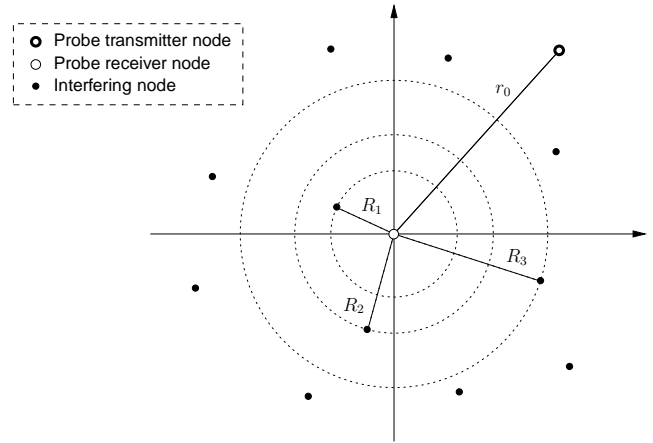}}
\par\end{centering}

\caption{\label{cap:2dplane}Poisson field model for the spatial distribution
of nodes. Without loss of generality, we assume the origin of the
coordinate system coincides with the probe receiver.}

\end{figure}

The proposed spatial model is depicted in Fig.~\ref{cap:2dplane}.
For analytical purposes, we assume there is a \emph{probe link} composed
of two nodes: the \emph{probe receiver}, located at the origin of
the two-dimensional plane (without loss of generality), and the \emph{probe
transmitter} (node~$\mbox{$i=0$}$), deterministically located at
a distance~$r_{0}$ from the origin. All the other nodes ($\mbox{$i=1\ldots\infty$}$)
are interfering nodes, whose random distances to the origin are denoted
by $\{R_{i}\}_{i=1}^{\infty}$, where $\mbox{$R_{1}\leq R_{2}\leq\ldots$}$.
Our goal is then to determine the effect of the interfering nodes
on the probe link.

\subsection{Transmission Characteristics of the Nodes}

To account for the transmission characteristics of users, we consider
that all interfering nodes employ the same linear modulation scheme,
such as $\mbox{$M$-ary}$ phase shift keying ($\mbox{$M$-PSK}$) or
$\mbox{$M$-ary}$ quadrature amplitude modulation ($\mbox{$M$-QAM}$),
with symbol period~$T$. Furthermore, they all transmit at the same
power~$P$ -- a plausible constraint when power control is too complex
to implement (e.g.,~decentralized ad-hoc networks). For generality,
however, we allow the probe transmitter to employ an arbitrary linear
modulation and arbitrary power~$P_{0}$, not necessarily equal to
those used by the interfering nodes. 

\begin{figure}
\begin{centering}
\scalebox{0.65}{\input{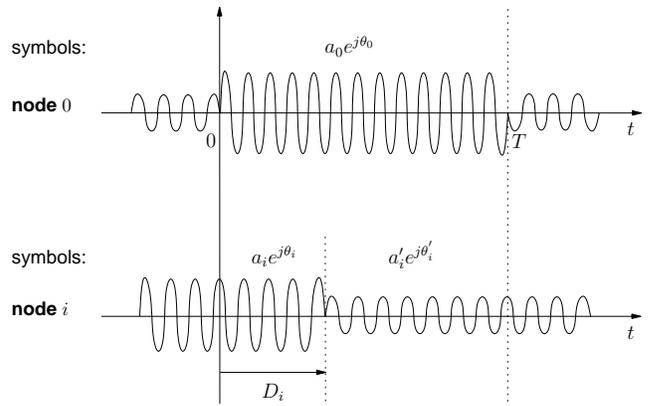}}
\par\end{centering}

\caption{\label{cap:sync}Asynchronism between different transmitting nodes.
In the observation interval~$[0,T]$, a change in constellation symbol
of node~$i$ occurs at random time~$\mbox{$t=D_{i}$}$, from $a_{i}e^{j\theta_{i}}$
to $a_{i}^{\prime}e^{j\theta_{i}^{\prime}}$, where $a$ and $\theta$
denote the transmitted symbol amplitude and phase, respectively. The
distribution of $D_{i}$ is assumed to be $\mathcal{U}(0,T)$. Therefore,
node~$0$ initiates symbol transmissions at times~$nT$ by convention,
while node~$i$ initiates symbol transmissions at times~$\mbox{$nT+D_{i}$}$.}

\end{figure}

We do not assume synchronization among interfering nodes, but instead
consider asynchronous transmissions where different terminals are
allowed to operate independently. As depicted in Fig.~\ref{cap:sync},
node~$i$ transmits with a random delay~$D_{i}$ relative to node~$0$,
where $\mbox{$D_{i}\sim\mathcal{U}(0,T)$}$.%
\footnote{We use $\mathcal{U}(a,b)$ to denote a real uniform distribution in
the interval~$[a,b]$.%
} The probe receiver employs a conventional linear detector.%
\footnote{Note that the other receiver nodes are not relevant for the analysis,
since they do not cause interference.%
} Typically, parameters such as the spatial density of interferers
and the propagation characteristics of the medium (e.g.,~shadowing
and path loss parameters) are unknown to the receiver. This lack of
information about the interference, together with constraints on receiver
complexity, justify the use of a simple linear detector, which is
optimal when only AWGN is present.

\subsection{Propagation Characteristics of the Medium}

To account for the propagation characteristics of the environment,
we consider that the median of the signal amplitude decays with the
distance~$r$ according to $k/r^{b}$, for some given constant~$k$.
The amplitude loss exponent~$b$ is environment-dependent, and can
approximately range from $0.8$ (e.g.,~hallways inside buildings)
to $4$ (e.g.,~dense urban environments), where $b=1$ corresponds
to free space propagation~\cite{Gol:05}.%
\footnote{Note that the \emph{amplitude} loss exponent is $b$, while the corresponding
\emph{power} loss exponent is $2b$.%
} The use of such a decay law also ensures that interferers located
far away have a negligible contribution to the total interference
observed at the probe receiver, thus making the infinite-plane model
reasonable.

To capture the shadowing effect, we model the channel amplitude gain~$S$
as a log-normal random variable (r.v.) such that $\mbox{$S=\mu e^{\sigma G}$}$,
where $\mbox{$G\sim\mathcal{N}(0,1)$}$,%
\footnote{We use $\mathcal{N}(\mu,\sigma^{2})$ to denote a real Gaussian distribution
with mean~$\mu$ and variance~$\sigma^{2}$.%
} $\mbox{$\mu=k/r^{b}$}$ is the median of $S$, and $\sigma$ is the
shadowing coefficient.%
\footnote{This model for combined path loss and log-normal shadowing can be
expressed in logarithmic form~\cite{Stu:00,Gol:05}, such that the
channel loss in dB is given by $L_{\textrm{dB}}=k_{0}+k_{1}\log_{10}r+\sigma_{\textrm{dB}}G$,
where $G\sim\mathcal{N}(0,1)$. The environment-dependent parameters~$(k_{0},k_{1},\sigma_{\textrm{dB}}$)
can be related to $(k,b,\sigma)$ as follows: $k_{0}=-20\log_{10}k$,
$k_{1}=20b$, and $\sigma_{\textrm{dB}}=\frac{20}{\ln10}\sigma$.
The parameter~$\sigma_{\textrm{dB}}$ is the standard deviation of
the channel loss in dB (or, equivalently, of the received SNR in dB),
and typically ranges from 6 to 12.%
} Thus, the shadowing is responsible for random fluctuations of the
channel gain around the median path gain~$k/r^{b}$. The multipath
effect is modeled as fast fading, which is superimposed on the path
loss and shadowing. Specifically, the fading affects the received
signal by introducing a random phase~$\mbox{$\phi\sim\mathcal{U}(0,2\pi)$}$,
as well as an amplitude factor~$\alpha$ with arbitrary distribution
and normalized to have unit power gain, i.e.,~$\mbox{$\mathbb{E}\{\alpha^{2}\}=1$}$.%
\footnote{We use $\mathbb{E}\{\cdot\}$ and $\mathbb{V}\{\cdot\}$ to denote
the expectation and variance operators, respectively.%
} Because of its fast nature, the fading is always averaged out in
this paper, both when determining the interference distribution and
the error probability. 

In what follows, we consider the shadowing (and similarly for the
fading) to be independent for different nodes~$i$, and approximately
constant during at least one symbol interval. Additionally, the probe
receiver can perfectly estimate the shadowing and fading affecting
its own link, hence ensuring that coherent demodulation of the desired
signal is possible.

\subsection{Mobility and Session Lifetime of the Interferers}

Typically, the time variation of the distances~$\{R_{i}\}_{i=1}^{\infty}$
of the interferers is highly coupled with that of the shadowing~$\{G_{i}\}_{i=1}^{\infty}$
affecting those nodes. This is because the shadowing is itself associated
with the movement of the nodes near large blocking objects. Thus,
we introduce the notation~$\mathcal{P}$ to denote \emph{{}``a particular
realization of the distances~$\{R_{i}\}_{i=1}^{\infty}$ and shadowing~$\{G_{i}\}_{i=1}^{\infty}$
of the interferers,''} or more succinctly, \emph{{}``the position
of the interferers}.'' In this paper, we analyze the following two
scenarios, which differ in the speed of variation of $\mathcal{P}$:
\begin{enumerate}
\item \emph{Slow-varying $\mathcal{P}$}: During the interval of interest
(e.g.,~a symbol or packet time), the distance~$R_{i}$ of each interferer
is approximately constant, $\mbox{$R_{i}(t)\approx R_{i}$}$. Furthermore,
the interferers have a long session lifetime, transmitting continuously
over many symbols. In this quasi-static scenario, $\mathcal{P}$ varies
slowly with time, and thus it is insightful to condition the interference
analysis on a given realization of $\mathcal{P}$. As we shall see,
this naturally leads to the derivation of the \emph{error outage probability}
of the probe link, which in this case is a more meaningful metric
than the error probability averaged over $\mathcal{P}$~\cite{AndTraVer:98,ConWinChi:J05,ConWinChi:J03,ConWinChiWint:L03}.
\item \emph{Fast-varying $\mathcal{P}$}: As in the previous case, $\mbox{$R_{i}(t)\approx R_{i}$}$
during the interval of interest. However, the interferers have a short
session lifetime, where each node periodically becomes active, transmits
a burst of symbols, and then turns off (e.g., in a sensor or a packet
network). Then, the set of \emph{interfering nodes} (the set of nodes
that are transmitting and contributing to the interference) changes
often, and so does their effective position~$\mathcal{P}$, which
experiences a variation analogous to that of a block fading model.
In this dynamic scenario, it is insightful to average the interference
analysis over all possible realizations of $\mathcal{P}$, which naturally
leads to the derivation of the \emph{average error probability} of
the probe link.
\end{enumerate}

\section{Interference Representation and Distribution\label{sec:Interference-Distribution}}

\subsection{Complex Baseband Representation of the Interference}

Under the system model described in Section~\ref{sec:System-model},
the aggregate signal~$Z(t)$ at the probe receiver can be written
for $\mbox{$0\leq t\leq T$}$ as\[
Z(t)=\frac{k\alpha_{0}e^{\sigma G_{0}}}{r_{0}^{b}}\sqrt{\frac{2}{T}}a_{0}\cos(2\pi f_{\mathrm{c}}t+\theta_{0})+Y(t)+W(t),\]
where the first right-hand term is the desired signal from the transmitter
probe node, $Y(t)$ is the aggregate interference with\begin{align*}
Y(t) & =\sum_{i=1}^{\infty}\frac{k\alpha_{i}e^{\sigma G_{i}}}{R_{i}^{b}}\left[\sqrt{\frac{2}{T}}a_{i}\cos(2\pi f_{\mathrm{c}}t+\theta_{i}+\phi_{i})u(D_{i}-t)\right.\\
 & \left.+\sqrt{\frac{2}{T}}a_{i}^{\prime}\cos(2\pi f_{\mathrm{c}}t+\theta_{i}^{\prime}+\phi_{i})u(t-D_{i})\right],\;0\leq t\leq T,\end{align*}
and $W(t)$ is the AWGN with two-sided power spectral density~$N_{0}/2$,
and independent of $Y(t)$. In the above equations, we use the following
the notation: $T$ is the symbol period; $f_{\mathrm{c}}$ is the
carrier frequency; $\mbox{$a_{i}e^{j\theta_{i}}$}$ and $\mbox{$a_{i}^{\prime}e^{j\theta_{i}^{\prime}}$}$
are r.v.'s denoting successive constellation symbols transmitted by
the node~$i$ during the interval of interest~$[0,T]$ (see Fig.~\ref{cap:sync});
and $u(t)$ is the unit step function. The overall effect of the path
loss, log-normal shadowing, and fading on node~$i$ is captured by
the amplitude factor~$k\alpha_{i}e^{\sigma G_{i}}/R_{i}^{b}$, where
$\mbox{$G_{i}\sim\mathcal{N}(0,1)$}$, and by the uniform phase~$\phi_{i}$.%
\footnote{Since we assume the probe receiver perfectly estimates the phase~$\phi_{0}$
of the multipath fading affecting its own link, we can set $\mbox{$\phi_{0}=0$}$
without loss of generality.%
} We consider that r.v.'s~$\alpha_{i}$, $\phi_{i}$, $G_{i}$, $R_{i}$,
$\mbox{$a_{i}e^{j\theta_{i}}$}$, $\mbox{$a_{i}^{\prime}e^{j\theta_{i}^{\prime}}$}$,
and $D_{i}$ are mutually independent for a given node~$i$, and
that the sequences $\{\alpha_{i}\}$, $\{\phi_{i}\}$, $\{G_{i}\}$,
$\{\mbox{$a_{i}e^{j\theta_{i}}$}\}$, $\{\mbox{$a_{i}^{\prime}e^{j\theta_{i}^{\prime}}$}\}$,
and $\{D_{i}\}$ are independent identically distributed (i.i.d.)
in $i$.

The probe receiver demodulates the desired signal from the aggregate
signal~$Z(t)$, using a conventional linear detector. This can be
achieved by projecting $Z(t)$ onto the orthonormal set~$\left\{ \mbox{$\psi_{1}(t)=\sqrt{2/T}\cos(2\pi f_{\mathrm{c}}t)$},\right.$
$\left.\mbox{$\psi_{2}(t)=-\sqrt{2/T}\sin(2\pi f_{\mathrm{c}}t)$}\right\} $.
Defining the in-phase and quadrature (IQ) components~$\mbox{$Z_{n}=\int_{0}^{T}Z(t)\psi_{n}(t)dt$}$,
$\mbox{$n=1,2$}$, we can write\begin{align}
Z_{1} & =\frac{k\alpha_{0}e^{\sigma G_{0}}}{r_{0}^{b}}a_{0}\cos\theta_{0}+Y_{1}+W_{1}\label{eq:z1}\\
Z_{2} & =\frac{k\alpha_{0}e^{\sigma G_{0}}}{r_{0}^{b}}a_{0}\sin\theta_{0}+Y_{2}+W_{2},\end{align}
where $W_{1}$ and $W_{2}$ are $\mathcal{N}(0,N_{0}/2)$ and mutually
independent. After some algebra (Appendix~\ref{sec:Deriv-BB-representation}),
$Y_{1}$ and $Y_{2}$ can be expressed as\begin{equation}
Y_{n}=\int_{0}^{T}Y(t)\psi_{n}(t)dt=\sum_{i=1}^{\infty}\frac{e^{\sigma G_{i}}X_{i,n}}{R_{i}^{b}},\quad n=1,2,\label{eq:Y1-sum}\end{equation}
 where\begin{align}
{\textstyle X_{i,1}} & {\textstyle =k\alpha_{i}\left[a_{i}\frac{D_{i}}{T}\cos(\theta_{i}+\phi_{i})+a_{i}^{\prime}\left(1-\frac{D_{i}}{T}\right)\cos(\theta_{i}^{\prime}+\phi_{i})\right]}\label{eq:Xi1}\\
{\textstyle X_{i,2}} & {\textstyle =k\alpha_{i}\left[a_{i}\frac{D_{i}}{T}\sin(\theta_{i}+\phi_{i})+a_{i}^{\prime}\left(1-\frac{D_{i}}{T}\right)\sin(\theta_{i}^{\prime}+\phi_{i})\right].}\label{eq:Xi2}\end{align}
 Using complex baseband notation,%
\footnote{Boldface letters are used to denote complex quantities; for example,
$\mbox{$\mathbf{Z}=Z_{1}+jZ_{2}$}$.%
} equations $\mbox{(\ref{eq:z1})-(\ref{eq:Xi2})}$ can be further simplified
as\begin{gather}
\mathbf{Z}=\frac{k\alpha_{0}e^{\sigma G_{0}}}{r_{0}^{b}}a_{0}e^{j\theta_{0}}+\mathbf{Y}+\mathbf{W}\label{eq:z}\\
\mathbf{Y}=\sum_{i=1}^{\infty}\frac{e^{\sigma G_{i}}\mathbf{X}_{i}}{R_{i}^{b}}\label{eq:ysum}\end{gather}
where\begin{equation}
\mathbf{X}_{i}=k\alpha_{i}e^{j\phi_{i}}\left[\frac{D_{i}}{T}a_{i}e^{j\theta_{i}}+\left(1-\frac{D_{i}}{T}\right)a_{i}^{\prime}e^{j\theta_{i}^{\prime}}\right],\label{eq:xi}\end{equation}
and the distribution of $\mathbf{W}$ is given by%
\footnote{We use $\mathcal{N}_{\textrm{c}}(0,\sigma^{2})$ to denote a circularly
symmetric (CS) complex Gaussian distribution, where the real and imaginary
parts are i.i.d.\ $\mathcal{N}(0,\sigma^{2}/2)$.%
}\begin{equation}
\mathbf{W}\sim\mathcal{N}_{\textrm{c}}(0,N_{0}).\label{eq:w-normal}\end{equation}
Since different interferers~$i$ transmit asynchronously and independently,
the r.v.'s~$\{\mathbf{X}_{i}\}_{i=1}^{\infty}$ are i.i.d. 

The distribution of the aggregate interference~$\mathbf{Y}$ plays
an important role in the evaluation of the error probability of the
probe link. In what follows, we characterize such distribution in
two important scenarios: the $\mbox{$\mathcal{P}$-conditioned}$ and
unconditional cases.

\subsection{$\mbox{$\mathcal{P}$-conditioned}$ Interference Distribution\label{sec:P-conditioned interference distribution}}

To derive the $\mbox{$\mathcal{P}$-conditioned}$ distribution of
the aggregate interference~$\mathbf{Y}$ in $\mbox{(\ref{eq:ysum})-(\ref{eq:xi})}$,
we start with the results given in \cite{GioChi:05}. This work shows
that in the case of Rayleigh fading, an expression of the form of
(\ref{eq:xi}) can be well approximated by a circularly symmetric
(CS) complex Gaussian r.v., such that\begin{equation}
\mathbf{X}_{i}\sim\mathcal{N}_{\textrm{c}}(0,2V_{X}),\quad V_{X}\triangleq\mathbb{V}\{X_{i,n}\}.\label{eq:Xi-gaussian-approx}\end{equation}
In \cite{GioChi:05}, the validity of this approximation is justified
both by analyzing the Kullback-Leibler divergence and comparing the
error probabilities in the exact and approximated cases.%
\footnote{We can obtain (\ref{eq:Xi-gaussian-approx}) following another approach:
if we consider that the interfering nodes are coded and operating
close to capacity, then the signal transmitted by each interferer
is Gaussian, such that $\mathbf{X}_{i}\sim\mathcal{N}_{\textrm{c}}(0,2V_{X})$~\cite{Fos:07}.%
} Then, conditioned on $\mathcal{P}$, the interference $\mbox{$\mathbf{Y}=\sum_{i=1}^{\infty}\frac{e^{\sigma G_{i}}\mathbf{X}_{i}}{R_{i}^{b}}$}$
becomes a sum of independent CS Gaussian r.v.'s and is therefore a
CS Gaussian r.v.\ given by%
\footnote{We use $X\,\mathop\sim\limits ^{{\scriptscriptstyle |Y}}$ to denote
the distribution of r.v.~$X$ conditional on $Y$.%
}\begin{equation}
\mathbf{Y}\,\mathop\sim\limits ^{{\scriptscriptstyle |\mathcal{P}}}\,\mathcal{N}_{\textrm{c}}(0,2AV_{X}),\label{eq:y-normal}\end{equation}
where $A$ is defined as\begin{equation}
A\triangleq\sum_{i=1}^{\infty}\frac{e^{2\sigma G_{i}}}{R_{i}^{2b}}.\label{eq:A}\end{equation}
Furthermore, we show in \cite{PinThe:06m} that after some algebra,
$V_{X}$ can be expressed as\begin{equation}
V_{X}=\frac{E}{3}+\frac{k^{2}}{6}\mathbb{E}\{a_{i}a_{i}^{\prime}\cos(\theta_{i}-\theta_{i}^{\prime})\},\quad i\geq1,\label{eq:Vx}\end{equation}
where $E\triangleq k^{2}\mathbb{E}\{a_{i}^{2}\}$ is the average symbol
energy of each interfering node, measured $1\,\textrm{m}$ away from
the interferer.%
\footnote{Unless otherwise stated, we will simply refer to $E$ as the {}``average
symbol energy'' of the interferers. %
} Because the r.v.'s~$\{\mathbf{X}_{i}\}_{i=1}^{\infty}$ are i.i.d.,
$V_{X}$ does not depend on $i$ and is only a function of the interferers'
signal constellation. For the case of equiprobable symbols and a constellation
that is symmetric with respect to the origin of the IQ-plane%
\footnote{A constellation is said to be \emph{symmetric with respect to the
origin} if for every constellation point~$\mbox{$(x,y)\in\mathbb{R}^{2}$}$,
the point~$(-x,-y)$ also belongs to the constellation. %
} (e.g.,~$\mbox{$M$-PSK}$ and $\mbox{$M$-QAM}$), the second right-hand
term in (\ref{eq:Vx}) vanishes and $\mbox{$V_{X}=E/3$}$.

\begin{figure}
\begin{centering}
\scalebox{1}{\psfrag{a}{\scriptsize{\sf{$a$}}}
\psfrag{fA(a)}{\scriptsize{\sf{$f_A(a)$}}}
\psfrag{bb=2, la=0.1}{\scriptsize{\sf{$b$=2, $\lambda$=0.1}}}
\psfrag{bb=2, la=0.15}{\scriptsize{\sf{$b$=2, $\lambda$=0.15}}}
\psfrag{bb=1.5, la=0.1}{\scriptsize{\sf{$b$=1.5, $\lambda$=0.1}}}
\psfrag{bb=1.5, la=0.15}{\scriptsize{\sf{$b$=1.5, $\lambda$=0.15}}}\includegraphics[width=1\columnwidth]{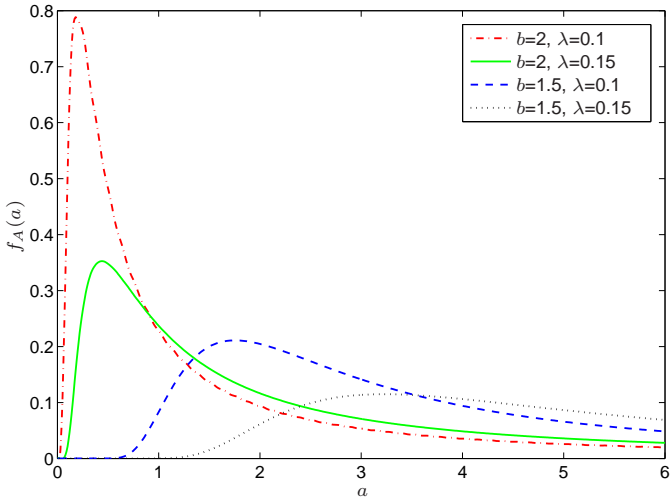}}
\par\end{centering}

\caption{\label{cap:pdf-a-m-lambda-plot}P.d.f.\ of $A$ for different amplitude
loss exponents~$b$ and interferer densities~$\lambda$ ($\sigma_{\textrm{dB}}=10$).
Stable laws are a direct generalization of Gaussian distributions,
and include other densities with heavier (algebraic) tails.}

\end{figure}

Lastly, note that since $A$ in (\ref{eq:A}) depends on the interferer
positions~$\mathcal{P}$ (i.e.,~$\{R_{i}\}_{i=1}^{\infty}$ and
$\{G_{i}\}_{i=1}^{\infty}$), it can be seen as a r.v.\ whose value
is different for each realization of $\mathcal{P}$. Furthermore,
Appendix~\ref{sec:Deriv-stable-A} shows that r.v.~$A$ has a \emph{skewed
stable distribution}~\cite{SamTaq:94} given by%
\footnote{We use $\mathcal{S}(\alpha,\beta,\gamma)$ to denote a real stable
distribution with characteristic exponent~$\mbox{$\alpha\in(0,2]$}$,
skewness~$\mbox{$\beta\in[-1,1]$}$, and dispersion~$\mbox{$\gamma\in[0,\infty)$}$.
The corresponding characteristic function is\[
\phi(w)=\begin{cases}
\exp\left[-\gamma|w|^{\alpha}\left(1-j\beta\mathrm{\, sign}(w)\tan\frac{\pi\alpha}{2}\right)\right], & \alpha\neq1,\\
\exp\left[-\gamma|w|\left(1+j\frac{2}{\pi}\beta\mathrm{\, sign}(w)\ln|w|\right)\right], & \alpha=1.\end{cases}\]
}\begin{equation}
A\sim\mathcal{S}\left(\alpha_{A}=\frac{1}{b},\:\beta_{A}=1,\:\gamma_{A}=\lambda\pi C_{1/b}^{-1}e^{2\sigma^{2}/b^{2}}\right),\label{eq:stableA}\end{equation}
where $\mbox{$b>1$}$, and $C_{x}$ is defined as\begin{equation}
C_{x}\triangleq\begin{cases}
\frac{1-x}{\Gamma(2-x)\cos(\pi x/2)}, & x\neq1,\\
\frac{2}{\pi}, & x=1.\end{cases}\label{eq:Cx}\end{equation}
This distribution is plotted in Fig.~\ref{cap:pdf-a-m-lambda-plot}
for different $b$ and $\lambda$.

\subsection{Unconditional Interference Distribution\label{sec:Unconditional-interference-distribution}}

To derive the unconditional distribution%
\footnote{Unconditional in the sense of being averaged over the positions~$\mathcal{P}$.%
} of the aggregate interference~$\mathbf{Y}$ in $\mbox{(\ref{eq:ysum})-(\ref{eq:xi})}$,
we can show that sums of the form of (\ref{eq:ysum}) belong to the
class of \emph{symmetric stable distributions}~\cite{SamTaq:94}.
This is because the r.v.'s~$\{R_{i}\}_{i=1}^{\infty}$ correspond
to distances in a spatial Poisson process and the $\{\mathbf{X}_{i}\}_{i=1}^{\infty}$
are i.i.d.\ and have a CS distribution. Specifically, Appendix~\ref{sec:Deriv-stable-Y}
shows that $\mathbf{Y}$ has a CS complex stable distribution given
by%
\footnote{We use $\mathcal{S}_{\textrm{c}}(\alpha,\beta=0,\gamma)$ to denote
a CS complex stable distribution with characteristic exponent~$\alpha$
and dispersion~$\gamma$, and whose characteristic function is $\phi(\mathbf{w})=\exp(-\gamma|\mathbf{w}|^{\alpha})$.
Furthermore, the corresponding real and imaginary components are both
$\mathcal{S}(\alpha,\beta=0,\gamma)$.%
}\begin{multline}
\mathbf{Y}\sim\mathcal{S}_{\textrm{c}}\left(\alpha_{\mathbf{Y}}=\frac{2}{b},\:\beta_{\mathbf{Y}}=0,\right.\\
\left.\gamma_{\mathbf{Y}}=\lambda\pi C_{2/b}^{-1}e^{2\sigma^{2}/b^{2}}\mathbb{E}\{|X_{i,n}|^{2/b}\}\right),\label{eq:stableY}\end{multline}
where $\mbox{$b>1$}$, and $C_{x}$ is defined in (\ref{eq:Cx}).
Using $\mbox{(\ref{eq:Xi1})-(\ref{eq:Xi2})}$, we can further express
$\mathbb{E}\{|X_{i,n}|^{2/b}\}$ in (\ref{eq:stableY}) as\begin{multline}
\mathbb{E}\{|X_{i,n}|^{2/b}\}=k^{2/b}\mathbb{E}\{|\alpha_{i}|^{2/b}\}\\
\times\underbrace{\mathbb{E}\left\{ \left|a_{i}\frac{D_{i}}{T}\cos(\theta_{i}+\phi_{i})+a_{i}^{\prime}\left(1-\frac{D_{i}}{T}\right)\cos(\theta_{i}^{\prime}+\phi_{i})\right|^{2/b}\right\} }_{\triangleq\,\chi(b)}.\label{eq:E-Xin}\end{multline}
For the particular case of Rayleigh fading, (\ref{eq:E-Xin}) reduces
to $\mathbb{E}\{|X_{i,n}|^{2/b}\}=k^{2/b}\Gamma\!\left(1+\frac{1}{b}\right)\cdot\chi(b),$
where we have used the moment relation for the Rayleigh r.v.'s~$\alpha_{i}$~\cite{Pro:00}.
Since different interferers~$i$ transmit asynchronously and independently,
the parameter~$\chi(b)$ does not depend on $i$ and is only a function
of the amplitude loss exponent~$b$ and the interferers' signal constellation.
Table~\ref{cap:table-E-Xij} provides some numerical values for $\mathbb{E}\{|X_{i,n}|^{2/b}\}$.

\begin{table}
\begin{onehalfspace}
\begin{centering}
\begin{tabular}{|c||c|c|}
\hline 
 & \multicolumn{2}{c|}{ $\frac{\mathbb{E}\{|X_{i,n}|^{2/b}\}}{E^{1/b}}$}\tabularnewline
$b$ & BPSK & QPSK\tabularnewline
\hline
\hline 
1.5 & $0.374$ & $0.385$\tabularnewline
\hline 
2 & $0.423$ & $0.441$\tabularnewline
\hline 
3 & $0.509$ & $0.531$\tabularnewline
\hline 
4 & $0.576$ & $0.599$\tabularnewline
\hline
\end{tabular}
\par\end{centering}
\end{onehalfspace}

\caption{\label{cap:table-E-Xij}$\mathbb{E}\{|X_{i,n}|^{2/b}\}$ for various
amplitude loss exponents~$b$ and modulations, assuming Rayleigh
fading. Note that for $\mbox{$M$-PSK}$ modulations, this quantity
is proportional to $E^{1/b}$, where $E$ is the average symbol energy
of the interferers.}

\end{table}

\subsection{Discussion\label{sec:interference-Discussion}}

The results of this section have to be interpreted carefully, because
of the different types of conditioning involved. In the unconditional
case, we let $\mathcal{P}$ be random, i.e.,~we let $\{R_{i}\}_{i=1}^{\infty}$
be the random outcomes of an underlying spatial Poisson process, and
$\{G_{i}\}_{i=1}^{\infty}$ be the random shadowing affecting each
interferer. Then, the unconditional interference~$\mathbf{Y}$ is
\emph{exactly} stable-distributed and given by (\ref{eq:stableY}).
We note that (\ref{eq:stableY}) and (\ref{eq:E-Xin}) hold for a
broad class of fading distributions, in addition to Rayleigh fading.
In the $\mbox{$\mathcal{P}$-conditioned}$ case, the positions of
the interferers are fixed. Then, $A$ in (\ref{eq:A}) is also a fixed
number, and the interference~$\mathbf{Y}$ is \emph{approximately}
CS Gaussian with total variance~$2AV_{X}$, as given in (\ref{eq:y-normal}).

\section{Error Probability\label{sec:Error-probability}}

In the previous section, we determined the statistical distribution
of the aggregate interference at the output of a conventional linear
receiver. We now use such result to directly characterize of the error
probability of the probe link, when subject to both interference and
thermal noise, in both cases of slow and fast-varying $\mathcal{P}$.

\subsection{Slow-varying Interferer Positions~$\mathcal{P}$\label{sec:Slow-varying-P}}

In the quasi-static scenario of slow-varying~$\mathcal{P}$, it is
insightful to analyze the error probability conditioned on a given
realization~$\mathcal{P}$ of the distances~$\{R_{i}\}_{i=1}^{\infty}$
and shadowing~$\{G_{i}\}_{i=1}^{\infty}$ associated with the interferers,
as well as on the shadowing~$G_{0}$ of the probe transmitter. We
denote this conditional symbol error probability by $P_{\textrm{e}}(G_{0},\mathcal{P})$.%
\footnote{The notation~$P_{\textrm{e}}(X,Y)$ is used as a shorthand for $\mathbb{P}\{\textrm{error}|X,Y$\}.%
}

To derive the conditional error probability, we employ the results
of Section~\ref{sec:P-conditioned interference distribution} for
the $\mbox{$\mathcal{P}$-conditioned}$ distribution of the aggregate
interference~$\mathbf{Y}$. Specifically, using (\ref{eq:w-normal})
and (\ref{eq:y-normal}), the received signal~$\mathbf{Z}$ in (\ref{eq:z})
can be rewritten as\begin{equation}
\mathbf{Z}=\frac{k\alpha_{0}e^{\sigma G_{0}}}{r_{0}^{b}}a_{0}e^{j\theta_{0}}+\mathbf{\widetilde{W}},\label{eq:Z-gaussian-approx}\end{equation}
where\begin{equation}
\mathbf{\widetilde{W}}=\mathbf{Y}+\mathbf{W}\,\mathop\sim\limits ^{{\scriptscriptstyle |\mathcal{P}}}\,\mathcal{N}_{\textrm{c}}(0,2AV_{X}+N_{0}),\label{eq:W'-gaussian-approx}\end{equation}
and $A$ was defined in (\ref{eq:A}). Our framework has thus reduced
the analysis to a Gaussian problem, where the combined noise~$\mathbf{\widetilde{W}}$
is Gaussian when conditioned on the location of the interferers.

\begin{figure}
\begin{centering}
\scalebox{0.8}{\input{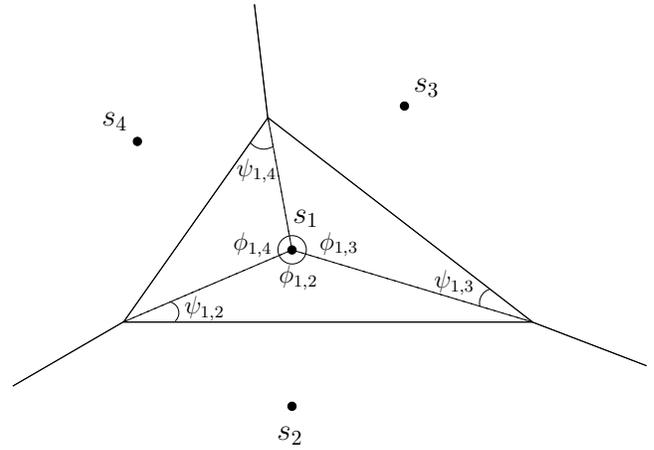}}
\par\end{centering}

\caption{\label{cap:decision-regions}Typical decision region associated with
symbol~$s_{1}$. In general, for a constellation with signal points~$\mbox{$s_{k}=|s_{k}|e^{j\xi_{k}}$}$
and $\mbox{$\zeta_{k}=\frac{|s_{k}|^{2}}{\mathbb{E}\{|s_{k}|^{2}\}}$}$,
$\mbox{$k=1\ldots M$}$, four parameters are required to compute the
error probability: $\phi_{k,l}$ and $\psi_{k,l}$ are the angles
that describe the decision region corresponding to $s_{k}$ (as depicted);
$\mathcal{B}_{k}$ is the set consisting of the indexes for the signal
points that share a decision boundary with $s_{k}$ (in the example,
$\mbox{$\mathcal{B}_{1}=\{2,3,4\}$}$); and $w_{k,l}=\zeta_{k}+\zeta_{l}-2\sqrt{\zeta_{k}\zeta_{l}}\cos(\xi_{k}-\xi_{l})$.}

\end{figure}

The corresponding error probability~$P_{\textrm{e}}(G_{0},\mathcal{P})$
can be found by taking the well-known error probability expressions
for coherent detection of linear modulations in the presence of AWGN
and fast fading~\cite{WinWint:J01,SimAlo:04,Cra:91,GifWinChi:J05},
but using $\mbox{$2AV_{X}+N_{0}$}$ instead of $N_{0}$ for the total
noise variance. Note that this substitution is valid for any linear
modulation, allowing the traditional results to be extended to include
the effect of network interference. For the case where the probe transmitter
employs an arbitrary signal constellation in the IQ-plane and the
fading is Rayleigh-distributed, the conditional symbol error probability~$P_{\textrm{e}}(G_{0},\mathcal{P})$
is given by\begin{multline}
P_{\textrm{e}}(G_{0},\mathcal{P})=\sum_{k=1}^{M}p_{k}\sum_{l\in\mathcal{B}_{k}}\frac{1}{2\pi}\\
\times\int_{0}^{\phi_{k,l}}\left(1+\frac{w_{k,l}}{4\sin^{2}(\theta+\psi_{k,l})}\eta_{A}\right)^{-1}d\theta,\label{eq:Pe-G0-P}\end{multline}
where\begin{equation}
\eta_{A}=\frac{e^{2\sigma G_{0}}E_{0}}{r_{0}^{2b}(2AV_{X}+N_{0})}\label{eq:eta-A}\end{equation}
is the received signal-to-interference-plus-noise ratio (SINR), averaged
over the fast fading; $M$ is the constellation size; $\{p_{k}\}_{k=1}^{M}$
are the symbol probabilities; $\mathcal{B}_{k}$, $\phi_{k,l}$, $w_{k,l}$,
and $\psi_{k,l}$ are the parameters that describe the geometry of
the constellation (see Fig.~\ref{cap:decision-regions}); $\mbox{$E_{0}\triangleq k^{2}\mathbb{E}\{a_{0}^{2}\}$}$
is the average symbol energy of the probe transmitter, measured $1\,\textrm{m}$
away from the transmitter; $A$ is defined in (\ref{eq:A}) and distributed
according to (\ref{eq:stableA}); and $V_{X}$ is given in (\ref{eq:Vx}).
When the probe transmitter employs $\mbox{$M$-PSK}$ and $\mbox{$M$-QAM}$
modulations with equiprobable symbols, (\ref{eq:Pe-G0-P}) is equivalent
to%
\footnote{In this paper, we implicitly assume that $\mbox{$M$-QAM}$ employs
a square signal constellation with $\mbox{$M=2^{n}$}$ points ($n$
even).%
}\begin{equation}
P_{\textrm{e}}^{\textrm{MPSK}}(G_{0},\mathcal{P})=\mathcal{I}_{A}\!\left(\frac{M-1}{M}\pi,\,\sin^{2}\left(\frac{\pi}{M}\right)\right)\label{eq:Pe-G0-P-PSK}\end{equation}
and\begin{align}
P_{\textrm{e}}^{\textrm{MQAM}}(G_{0},\mathcal{P}) & =4\left(1-\frac{1}{\sqrt{M}}\right)\cdot\mathcal{I}_{A}\!\left(\frac{\pi}{2},\,\frac{3}{2(M-1)}\right)\nonumber \\
 & -4\left(1-\frac{1}{\sqrt{M}}\right)^{2}\cdot\mathcal{I}_{A}\!\left(\frac{\pi}{4},\,\frac{3}{2(M-1)}\right),\label{eq:Pe-G0-P-QAM}\end{align}
respectively, where the integral~$\mathcal{I}_{A}(x,g)$ is given
by\begin{equation}
\mathcal{I}_{A}(x,g)=\frac{1}{\pi}\int_{0}^{x}\left(1+\frac{g}{\sin^{2}\theta}\eta_{A}\right)^{-1}d\theta.\label{eq:L-x-g}\end{equation}

In the general expression given in (\ref{eq:Pe-G0-P}) and (\ref{eq:eta-A}),
the network interference is accounted for by the term~$2AV_{X}$,
where $A$ depends on the spatial distribution of the interferers
and propagation characteristics of the medium, while $V_{X}$ depends
on the interferer transmission characteristics. Since $2AV_{X}$ simply
adds to $N_{0}$, we conclude that the effect of the interference
on the error probability is simply an increase in the noise level,
a fact which is intuitively satisfying. Furthermore, note that the
modulation of the interfering nodes only affects the term~$V_{X}$,
while the (possibly different) modulation of the probe transmitter
affects the \emph{type} of error probability expression, leading to
forms such as (\ref{eq:Pe-G0-P-PSK}) or (\ref{eq:Pe-G0-P-QAM}). 

In our quasi-static model, the conditional error probability in (\ref{eq:Pe-G0-P})
is seen to be a function of the slow-varying user positions and shadowing
(i.e.,~$G_{0}$ and $\mathcal{P}$). Since these quantities are random,
the error probability itself is a r.v. Then, with some probability,
$G_{0}$ and $\mathcal{P}$ are such that the error probability of
the probe link is above some target~$p^{*}$. The system is said
to be \emph{in outage}, and the error outage probability is\begin{equation}
P_{\textrm{out}}^{\textrm{e}}=\mathbb{P}_{G_{0},\mathcal{P}}\{P_{\textrm{e}}(G_{0},\mathcal{P})>p^{*}\},\label{eq:Pout-slowP}\end{equation}
In the case of slow-varying user positions, the error outage probability
is a more meaningful metric than the error probability averaged over
$\mathcal{P}$.

\subsection{Fast-varying Interferer Positions~$\mathcal{P}$\label{sec:fast-varying-P}}

In the dynamic scenario of fast-varying $\mathcal{P}$, it is insightful
to average the error probability over all possible realizations of
interferer positions~$\mathcal{P}$. We denote this average symbol
error probability by $P_{\textrm{e}}(G_{0})$. Note that we choose
not to average out the shadowing~$G_{0}$ affecting the probe transmitter,
since we have assumed the probe transmitter node is immobile at a
deterministic distance~$r_{0}$ from the probe receiver, and thus
$G_{0}$ is slow-varying.

To derive the average error probability, we use the decomposition
property of stable r.v.'s~\cite{SamTaq:94}, which allows $\mathbf{Y}$
in (\ref{eq:stableY}) to be decomposed as\begin{equation}
\mathbf{Y}=\sqrt{B}\mathbf{G},\label{eq:y-decomposition}\end{equation}
where $B$ and $\mathbf{G}$ are independent r.v.'s, and\begin{gather}
B\sim\mathcal{S}\left(\alpha_{B}=\frac{1}{b},\:\beta_{B}=1,\:\gamma_{B}=\cos\frac{\pi}{2b}\right)\label{eq:stableB}\\
\mathbf{G}\sim\mathcal{N}_{\textrm{c}}(0,2V_{G}),\; V_{G}=2e^{2\sigma^{2}/b}\left(\lambda\pi C_{2/b}^{-1}\mathbb{E}\{|X_{i,n}|^{2/b}\}\right)^{b},\label{eq:G-and-Vg}\end{gather}
 with $\mathbb{E}\{|X_{i,n}|^{2/b}\}$ given in (\ref{eq:E-Xin}).
Conditioning on the r.v.~$B$, we then use (\ref{eq:w-normal}) and
(\ref{eq:y-decomposition}) to rewrite the aggregate received signal~$\mathbf{Z}$
in (\ref{eq:z}) as\[
\mathbf{Z}=\frac{k\alpha_{0}e^{\sigma G_{0}}}{r_{0}^{b}}a_{0}e^{j\theta_{0}}+\mathbf{\widetilde{W}},\]
where\begin{equation}
\mathbf{\widetilde{W}}=\sqrt{B}\mathbf{G}+\mathbf{W}\,\mathop\sim\limits ^{{\scriptscriptstyle |B}}\,\mathcal{N}_{\textrm{c}}(0,2BV_{G}+N_{0}).\label{eq:W'-normal}\end{equation}
Again, our framework has reduced the analysis to a Gaussian problem,
where the combined noise~$\mathbf{\widetilde{W}}$ is a Gaussian
r.v. Note that this result was derived without resorting to any approximations
-- in particular, the Gaussian approximation of (\ref{eq:Xi-gaussian-approx})
was not needed here. We merely used the decomposition property of
symmetric stable r.v.'s. 

The corresponding error probability~$P_{\textrm{e}}(G_{0})$ can
be found by taking the error expressions for coherent detection in
the presence of AWGN and fast fading, then using $\mbox{$2BV_{G}+N_{0}$}$
instead of $N_{0}$ for the total noise variance, and lastly (unlike
in Section~\ref{sec:Slow-varying-P}) averaging over the r.v.~$B$.
For the case where the probe transmitter employs an arbitrary signal
constellation in the IQ-plane and the fading is Rayleigh-distributed,
the average symbol error probability~$P_{\textrm{e}}(G_{0})$ is
given by\begin{multline}
P_{\textrm{e}}(G_{0})=\sum_{k=1}^{M}p_{k}\sum_{l\in\mathcal{B}_{k}}\frac{1}{2\pi}\\
\times\int_{0}^{\phi_{k,l}}\mathbb{E}_{B}\left\{ \left(1+\frac{w_{k,l}}{4\sin^{2}(\theta+\psi_{k,l})}\eta_{B}\right)^{-1}\right\} d\theta,\label{eq:Pe-G0}\end{multline}
where\begin{equation}
\eta_{B}=\frac{e^{2\sigma G_{0}}E_{0}}{r_{0}^{2b}(2BV_{G}+N_{0})};\label{eq:eta-B}\end{equation}
$B$ is distributed according to (\ref{eq:stableB}); $V_{G}$ is
given in (\ref{eq:G-and-Vg}); and the other parameters have the same
meaning as in Section~\ref{sec:Slow-varying-P}. When the probe transmitter
employs $\mbox{$M$-PSK}$ and $\mbox{$M$-QAM}$ modulations with equiprobable
symbols, (\ref{eq:Pe-G0-P}) is equivalent to\begin{equation}
{\textstyle P_{\textrm{e}}^{\textrm{MPSK}}(G_{0})=\mathcal{I}_{B}\!\left(\frac{M-1}{M}\pi,\,\sin^{2}\left(\frac{\pi}{M}\right)\right)}\label{eq:Pe-G0-PSK}\end{equation}
and\begin{align}
P_{\textrm{e}}^{\textrm{MQAM}}(G_{0}) & =4\left(1-\frac{1}{\sqrt{M}}\right)\cdot\mathcal{I}_{B}\!\left(\frac{\pi}{2},\,\frac{3}{2(M-1)}\right)\nonumber \\
 & -4\left(1-\frac{1}{\sqrt{M}}\right)^{2}\cdot\mathcal{I}_{B}\!\left(\frac{\pi}{4},\,\frac{3}{2(M-1)}\right),\label{eq:Pe-G0-QAM}\end{align}
respectively, where the integral~$\mathcal{I}_{B}(x,g)$ is given
by\begin{equation}
\mathcal{I}_{B}(x,g)=\frac{1}{\pi}\int_{0}^{x}\mathbb{E}_{B}\left\{ \left(1+\frac{g}{\sin^{2}\theta}\eta_{B}\right)^{-1}\right\} d\theta.\label{eq:D-x-g}\end{equation}

\subsection{Discussion}

\begin{figure}
\begin{centering}
\scalebox{0.45}{\psfrag{INR (dB)}{\Large{$\mathsf{INR\;(dB)}$}}
\psfrag{interferer density la (m2)}{\Large{\sf{interferer density $\lambda$ ($\mathsf{m^{-2}}$)}}}
\psfrag{Peout=5e-3,e-2,5e-2}{\LARGE{$P_{\textrm{out}}^{\textrm{e}}=\mathsf{5\cdot10^{-3},10^{-2},5\cdot10^{-2}}$}}\includegraphics{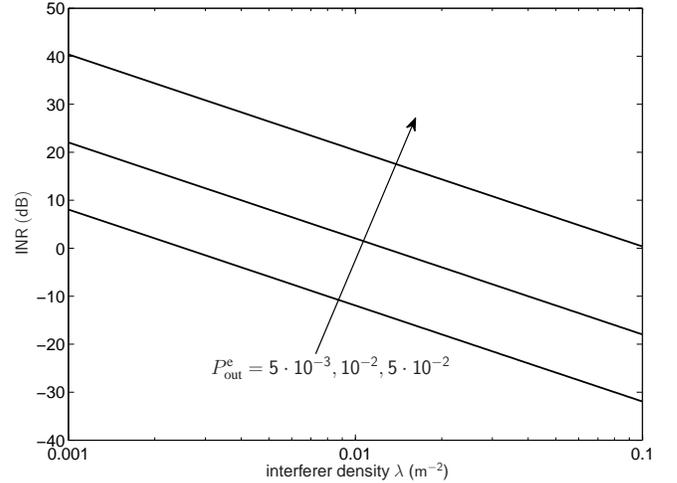}}
\par\end{centering}

\caption{\label{cap:const_lambda_inr_plot}$\mbox{$\mathsf{INR}-\lambda$}$
curves of constant $P_{\textrm{out}}^{\textrm{e}}$, for the case
of slow-varying interferer positions~$\mathcal{P}$ (BPSK, $\mbox{$\mathsf{SNR}=40\,\textrm{dB}$}$,
$\mbox{$b=2$}$, $\mbox{$r_{0}=1\,\textrm{m}$}$, $\sigma_{\textrm{dB}}=10$,
$\mbox{$p^{*}=10^{-2}$}$). INR is the interference-to-noise ratio,
defined as $\mbox{$\mathsf{INR}=E/N_{0}$}$. Clearly, for a fixed
error performance, there is a tradeoff between the density and energy
of the interferers: if the INR (or, equivalently, $E$) increases,
$\lambda$ must decrease, and vice-versa, to maintain the same outage
probability.}

\end{figure}

Using the results derived in Sections~\ref{sec:Slow-varying-P} and
\ref{sec:fast-varying-P}, we can now analyze the dependence of the
error performance on the density~$\lambda$ and the average  symbol
energy~$E$ of the interfering nodes. For that purpose, we use (\ref{eq:Pe-G0-P}),
although (\ref{eq:Pe-G0}) would lead to similar conclusions. In (\ref{eq:Pe-G0-P}),
the error probability~$P_{\textrm{e}}(G_{0},\mathcal{P})$ implicitly
depends on parameters $\lambda$ and $E$ through the product~$AV_{X}$
in the denominator of $\eta_{A}$ in (\ref{eq:eta-A}). This is because
the dispersion parameter~$\gamma_{A}$ of the stable r.v.~$A$ depends
on $\lambda$ according to (\ref{eq:stableA}), and $V_{X}$ is proportional
to $E$ as in (\ref{eq:Vx}). The dependence on $\lambda$ can be
made evident by using the scaling property of stable r.v.'s~\cite{SamTaq:94}
to write $\mbox{$AV_{X}=\lambda^{b}\widetilde{A}V_{X}$}$, where $\widetilde{A}$
is a normalized version of $A$, independent of $\lambda$. We thus
conclude that the interference term~$AV_{X}$ is proportional to
$\lambda^{b}E$, where $\mbox{$b>1$}$. Clearly, the error performance
degrades faster with an increase in the \emph{density} of interferers
than with an increase in their \emph{transmitted power}. The tradeoff
between $E$ and $\lambda$ for a fixed error performance is illustrated
in Fig.~\ref{cap:const_lambda_inr_plot}.

\subsection{Numerical Results\label{sec:error-plots}}

\begin{figure}
\begin{centering}
\subfigure[$P_{\textrm{out}}^{\textrm{e}}$ versus the SNR of the probe link, for various interference-to-noise ratios~INR (BPSK, $\mbox{$b=2$}$, $\mbox{$\lambda=0.01\,\textrm{m}^{-2}$}$, $\mbox{$r_{0}=1\,\textrm{m}$}$, $\mbox{$\sigma_{\textrm{dB}}=10$}$, $\mbox{$p^{*}=10^{-2}$}$).]{
  \label{cap:pout-approx-inr-plot}
  \scalebox{0.45}{
    \psfrag{Peout}{\LARGE{$P_{\textrm{out}}^{\textrm{e}}$}}
    \psfrag{SNR (dB)}{\Large{$\mathsf{SNR\;(dB)}$}}
    \psfrag{INR=-inf,10,20,30 dB}{\LARGE{$\mathsf{INR=-\infty,10,20,30\;dB}$}}
    \includegraphics{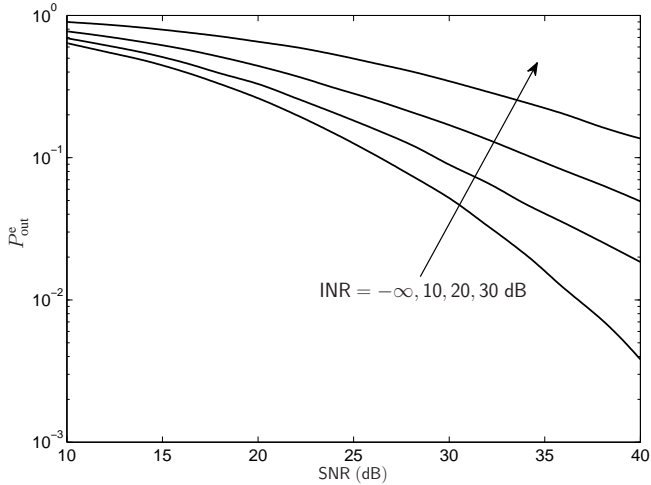}
  }
}
\par\end{centering}

\begin{centering}
\subfigure[$P_{\textrm{out}}^{\textrm{e}}$ versus the SNR of the probe link, for various interferer spatial densities~$\lambda$ (BPSK, $\mbox{$\mathsf{INR}=10\,\textrm{dB}$}$, $\mbox{$b=2$}$, $\mbox{$r_{0}=1\,\textrm{m}$}$, $\mbox{$\sigma_{\textrm{dB}}=10$}$, $\mbox{$p^{*}=10^{-2}$}$).]{
   \label{cap:pout-approx-lambda-plot}
   \scalebox{0.45}{
    \psfrag{Peout}{\LARGE{$P_{\textrm{out}}^{\textrm{e}}$}}
    \psfrag{SNR (dB)}{\Large{$\mathsf{SNR\;(dB)}$}}
    \psfrag{la=0,0.01,0.1,1}{\LARGE{$\mathsf{\lambda=0,0.01,0.1,1\, m^{-2}}$}}
    \includegraphics{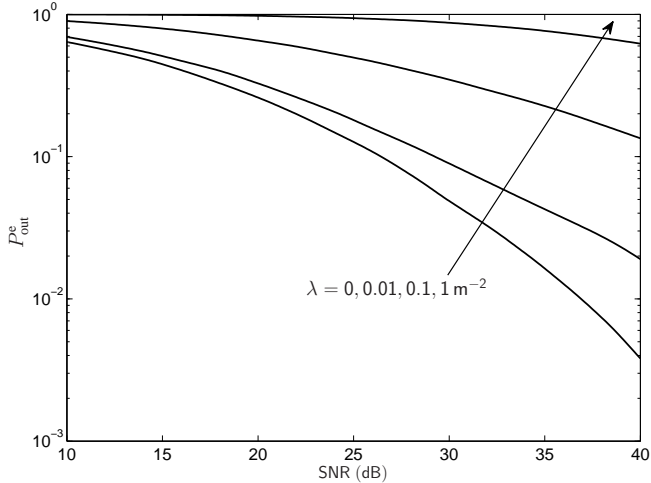}
  }
}
\par\end{centering}

\caption{\label{cap:hetero-slow-P}Error outage probability plots for a heterogeneous
network (where $\mbox{$\mathsf{SNR}\neq\mathsf{INR}$}$ in general)
and slow-varying interferer positions~$\mathcal{P}$. Since $\mathcal{P}$
is slow-varying, the meaningful performance metric is the outage probability~$P_{\textrm{out}}^{\textrm{e}}$
given in (\ref{eq:Pout-slowP}).}

\end{figure}

\begin{figure}
\begin{centering}
\subfigure[$P_{\textrm{e}}(G_{0})$ versus the length~$r_{0}$ of the probe link, for various signal loss exponents~$b$ (BPSK, $\mbox{$G_{0}=0$}$, $\mbox{$\mathsf{SNR}=\mathsf{INR}=20\,\textrm{dB}$}$, $\mbox{$\lambda=0.01\,\textrm{m}^{-2}$}$, $\mbox{$\sigma_{\textrm{dB}}=10$}$).]{
  \label{cap:pe-exact-m-plot}
  \scalebox{0.45}{
    \psfrag{Pe(G0)}{\LARGE{$P_{\textrm{e}}(G_{0})$}}
    \psfrag{probe link length r0 (m)}{\Large{\sf{probe link length $r_0$ (m)}}}
    \psfrag{b=4,2,1.5}{\LARGE{$b=\mathsf{4,2,1.5}$}}
    \includegraphics{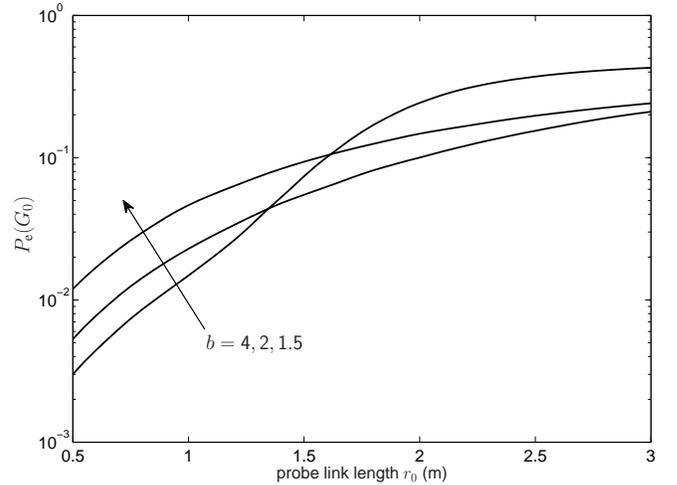}
  }
}
\par\end{centering}

\begin{centering}
\subfigure[$P_{\textrm{e}}(G_{0})$ versus the SNR, for various interferer densities~$\lambda$ (BPSK, $\mbox{$G_{0}=0$}$, $\mbox{$b=3$}$, $\mbox{$r_{0}=1\,\textrm{m}$}$, $\mbox{$\sigma_{\textrm{dB}}=10$}$).]{
   \label{cap:pe-exact-snr-inr-lambda-plot}
   \scalebox{0.45}{
     \psfrag{Pe(G0)}{\LARGE{$P_{\textrm{e}}(G_{0})$}}
     \psfrag{SNR (dB)}{\Large{$\mathsf{SNR\;(dB)}$}}
     \psfrag{la=0,10e-3,10e-2,10e-1}{\LARGE{$\mathsf{\lambda=0,10^{-3},10^{-2},10^{-1}\, m^{-2}}$}}
     \includegraphics{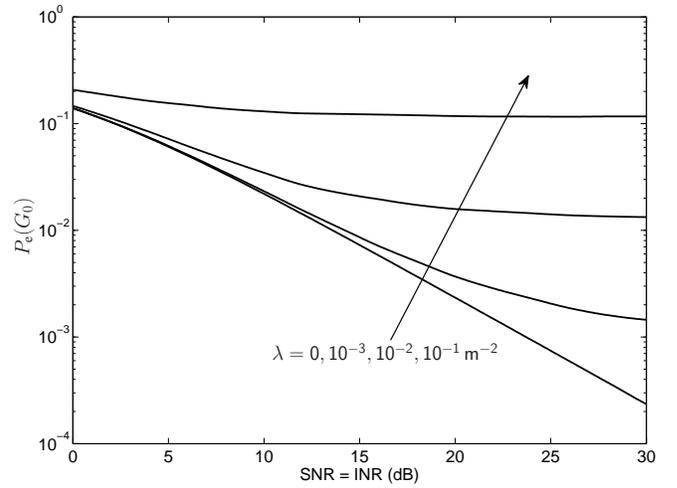}
  }
}
\par\end{centering}

\caption{\label{cap:homo-fast-P}Average error probability plots for a homogeneous
network (where $\mbox{$\mathsf{SNR}=\mathsf{INR}$}$) and fast-varying
interferer positions~$\mathcal{P}$. Since $\mathcal{P}$ is fast-varying,
the meaningful performance metric is the average error probability~$P_{\textrm{e}}(G_{0})$
given in (\ref{eq:Pe-G0}). For simplicity, we use $\mbox{$G_{0}=0$}$
in these plots (no shadowing on the probe link).}

\end{figure}

Figs.~\ref{cap:hetero-slow-P} and \ref{cap:homo-fast-P} quantify
the average and outage probabilities for several scenarios, showing
their dependence on various parameters involved, such as the signal-to-noise
ratio~$\mbox{$\mathsf{SNR}=E_{0}/N_{0}$}$, interference-to-noise
ratio~$\mbox{$\mathsf{INR}=E/N_{0}$}$, amplitude loss exponent~$b$,
interferer spatial density~$\lambda$, and link length~$r_{0}$.

The plots of $P_{\textrm{out}}^{\textrm{e}}$ and $P_{\textrm{e}}(G_{0})$
presented here are of semi-analytical nature. Specifically, we resort
to a hybrid method where we employ the analytical results given in
$\mbox{(\ref{eq:Pe-G0-P})-(\ref{eq:Pout-slowP})}$ and $\mbox{(\ref{eq:Pe-G0})-(\ref{eq:D-x-g})}$,
and perform a Monte Carlo simulation with respect to the stable r.v.'s
(i.e.,~$A$ and $B$), according to \cite{ChaMalStu:76}. Nevertheless,
we emphasize that the expressions derived in this paper completely
eliminate the need for simulation of the interferers' position and
waveforms in the network, in order to obtain the error performance.

For illustration purposes, the plots assume that all terminals (i.e.,~the
probe transmitter and interfering nodes) use BPSK modulation. We analyze
both cases of slow and fast-varying interferer positions~$\mathcal{P}$,
concurrently with the following two different scenarios:
\begin{enumerate}
\item \emph{Heterogeneous network}: The probe transmitter is allowed to
use an arbitrary power~$P_{0}=E_{0}/T$, not necessarily equal to
the common power of the interfering nodes~$P=E/T$, and hence $\mbox{$\mathsf{SNR}\neq\mathsf{INR}$}$
in general. This scenario is useful when the goal is to evaluate the
impact of aggregate interference from a large number of identical
secondary users (e.g.,~cognitive-radio terminals) on the performance
of a primary link.
\item \emph{Homogeneous network}: The probe transmitter and interfering
nodes all use the same power, and thus $\mbox{$\mathsf{SNR}=\mathsf{INR}$}$.
This may correspond to a sensor network scenario, where there is a
large number of indistinguishable, spatially scattered nodes with
similar transmission characteristics. In such a case, the goal is
to evaluate the impact of the aggregate network self-interference
on the performance of each sensor node.
\end{enumerate}
For the heterogeneous case depicted in Fig.~\ref{cap:hetero-slow-P},
we conclude that the error performance deteriorates as $\lambda$
or INR increase, for a fixed SNR. This is expected because as the
density or transmitted energy of the interferers increase, the aggregate
interference at the probe receiver becomes stronger. Note, however,
that in the homogeneous case where $\mbox{$\mathsf{SNR}=\mathsf{INR}$}$,
the error performance improves as we increase the common transmitted
power~$P$ of the nodes (or equivalently, the SNR), although the
gains become marginally small as $\mbox{$P\rightarrow\infty$}$ (see
Fig.~\ref{cap:pe-exact-snr-inr-lambda-plot}). This happens because
in the interference-limited regime where $\mbox{$\mathsf{SNR}=\mathsf{INR}\gg1$}$,
the noise term $N_{0}$ in (\ref{eq:eta-A}) or (\ref{eq:eta-B})
becomes irrelevant, and so the SNR in the numerator cancels with the
INR in the denominator, making the performance independent of the
transmitted power~$P$.

The effect of the amplitude loss exponent~$b$ on the error performance,
on the other hand, is non-trivial. As illustrated in Fig.~\ref{cap:pe-exact-m-plot},
an increase in $b$ may degrade or improve the performance, depending
on the value of the link length~$r_{0}$ and other parameters. This
is because $b$ simultaneously affects both the received signal of
interest and the aggregate interference -- the former, through the
term~$1/r_{0}^{b}$; and the latter, through $\alpha_{A}$ and $\gamma_{A}$
in (\ref{eq:stableA}), or through $\alpha_{B}$, $\gamma_{B}$, and
$V_{G}$ in (\ref{eq:stableB}) and (\ref{eq:G-and-Vg}).

\section{Summary\label{sec:Summary}}

This paper introduces a mathematical model for communication subject
to network interference and noise. The interferers are scattered according
to a spatial Poisson process, and are operating asynchronously in
a wireless environment subject to path loss, shadowing, and multipath
fading. We show that the aggregate network interference at the output
of a linear receiver is related to a \emph{skewed stable distribution}
when conditioned on the positions of interferers, and to a \emph{symmetric
stable distribution} in the unconditional case. We characterize the
error performance for the cases of slow and fast-varying interferers,
in terms of outage and average error probabilities, respectively.
These expressions are valid for any linear modulation scheme. We then
quantify these metrics as a function of various important system parameters,
such as the SNR, INR, path loss exponent, and spatial density of the
interferers. In Part~II of the paper~\cite{PinWin:J09a}, we characterize
the capacity of the link when subject to both network interference
and noise, and derive the spectrum of the aggregate interference at
any location in the plane. Lastly, we put forth the concept of spectral
outage probability, a new characterization of the aggregate interference
generated by communicating nodes in a wireless network.

\appendices

\section{Derivation of the Complex Baseband Interference Representation\label{sec:Deriv-BB-representation}}

To derive the representation (\ref{eq:ysum}) and (\ref{eq:xi}) of
the aggregate interference~$Y(t)$, we project $Y(t)$ onto the basis
function~$\mbox{$\psi_{1}(t)=\sqrt{2/T}\cos(2\pi f_{\mathrm{c}}t)$}$
as follows:\begin{align*}
Y_{1}= & \int_{0}^{T}Y(t)\psi_{1}(t)dt\\
= & \sum_{i=1}^{\infty}\int_{0}^{T}\frac{k\alpha_{i}e^{\sigma G_{i}}}{R_{i}^{b}}\left[\sqrt{\frac{2}{T}}a_{i}\cos(2\pi f_{\mathrm{c}}t+\theta_{i}+\phi_{i})u(D_{i}-t)\right.\\
 & \left.+\sqrt{\frac{2}{T}}a_{i}^{\prime}\cos(2\pi f_{\mathrm{c}}t+\theta_{i}^{\prime}+\phi_{i})u(t-D_{i})\right]\\
 & \times\sqrt{\frac{2}{T}}\cos(2\pi f_{\mathrm{c}}t)dt\\
= & \sum_{i=1}^{\infty}\frac{2}{T}\frac{k\alpha_{i}e^{\sigma G_{i}}}{R_{i}^{b}}\left[\frac{a_{i}}{2}\int_{0}^{D_{i}}\cos(\theta_{i}+\phi_{i})dt\right.\\
 & +\frac{a_{i}}{2}\underbrace{\int_{0}^{D_{i}}\cos(4\pi f_{\mathrm{c}}t+\theta_{i}+\phi_{i})dt}_{\approx0\;\textrm{for}\; f_{\mathrm{c}}T\gg1}+\frac{a_{i}^{\prime}}{2}\int_{D_{i}}^{T}\cos(\theta_{i}^{\prime}+\phi_{i})dt\\
 & \left.+\frac{a_{i}^{\prime}}{2}\underbrace{\int_{D_{i}}^{T}\cos(4\pi f_{\mathrm{c}}t+\theta_{i}^{\prime}+\phi_{i})dt}_{\approx0\;\textrm{for}\; f_{\mathrm{c}}T\gg1}\right]\\
 & =\sum_{i=1}^{\infty}\frac{e^{\sigma G_{i}}X_{i,1}}{R_{i}^{b}},\end{align*}
where\[
X_{i,1}=k\alpha_{i}\left[a_{i}\frac{D_{i}}{T}\cos(\theta_{i}+\phi_{i})+a_{i}^{\prime}\left(1-\frac{D_{i}}{T}\right)\cos(\theta_{i}^{\prime}+\phi_{i})\right].\]

The signal~$Y(t)$ can be projected onto the basis function~$\mbox{$\psi_{2}(t)=-\sqrt{2/T}\sin(2\pi f_{\mathrm{c}}t)$}$
in an entirely analogous way, leading to\[
Y_{2}=\sum_{i=1}^{\infty}\frac{e^{\sigma G_{i}}X_{i,2}}{R_{i}^{b}},\]
where\[
X_{i,2}=k\alpha_{i}\left[a_{i}\frac{D_{i}}{T}\sin(\theta_{i}+\phi_{i})+a_{i}^{\prime}\left(1-\frac{D_{i}}{T}\right)\sin(\theta_{i}^{\prime}+\phi_{i})\right].\]
We can combine $X_{i,1}$ and $X_{i,2}$ in the complex r.v.~$\mbox{$\mathbf{X}_{i}=X_{i,1}+jX_{i,2}$}$
as\[
\mathbf{X}_{i}=k\alpha_{i}e^{j\phi_{i}}\left[\frac{D_{i}}{T}a_{i}e^{j\theta_{i}}+\left(1-\frac{D_{i}}{T}\right)a_{i}^{\prime}e^{j\theta_{i}^{\prime}}\right],\]
which completes the derivation.

\section{Derivation of the Distribution of $A$\label{sec:Deriv-stable-A}}

To derive the distribution of $A$ given in (\ref{eq:stableA}), we
start with the following theorem.

\begin{theorem} Let $\{\tau_{i}\}_{i=1}^{\infty}$ denote the arrival
times of a one-dimensional Poisson process with rate~$\lambda$;
let $\{W_{i}\}_{i=1}^{\infty}$ be a sequence of nonnegative i.i.d.\ r.v.'s,
independent of the sequence~$\{\tau_{i}\}$ and satisfying $\mbox{$\mathbb{E}\{|W_{i}|^{\alpha}\}<\infty$}$.
If $\mbox{$0<\alpha<1$}$, then\[
\sum_{i=1}^{\infty}\frac{W_{i}}{\tau_{i}^{1/\alpha}}\,\mathop\sim\limits ^{{\scriptscriptstyle \mathrm{a.s.}}}\,\mathcal{S}\left(\alpha,\:\beta=1,\:\gamma=\lambda C_{\alpha}^{-1}\mathbb{E}\{|W_{i}|^{\alpha}\}\right),\]
where $C_{\alpha}$ is defined in (\ref{eq:Cx}). 

\end{theorem}

\begin{proof}See \cite{SamTaq:94}.\end{proof}

If an homogeneous Poisson point process \emph{in the plane} has spatial
density~$\lambda$, and $R_{i}$ denotes the distance of node~$i$
to the origin, then the sequence~$\{R_{i}^{2}\}_{i=1}^{\infty}$
represents Poisson arrival times \emph{on the line} with the constant
arrival rate~$\lambda\pi$. This can be easily shown by mapping the
spatial Poisson process from Cartesian into polar coordinates, and
then applying the mapping theorem~\cite{Kin:93}. Using this fact,
we can then apply the above theorem to (\ref{eq:A}) and write\begin{multline}
A=\sum_{i=1}^{\infty}\frac{e^{2\sigma G_{i}}}{R_{i}^{2b}}=\sum_{i=1}^{\infty}\frac{\overbrace{e^{2\sigma G_{i}}}^{W_{i}}}{(\underbrace{R_{i}^{2}}_{\tau_{i}})^{b}}\\
\,\mathop\sim\limits ^{{\scriptscriptstyle \mathrm{a.s.}}}\,\mathcal{S}\left(\alpha=\frac{1}{b},\:\beta=1,\:\gamma=\lambda\pi C_{1/b}^{-1}\mathbb{E}\{|e^{2\sigma G_{i}}|^{1/b}\}\right),\label{eq:A-step1}\end{multline}
for $\mbox{$b>1$}$. Using the moment property of log-normal r.v.'s,
i.e., $\mbox{$\mathbb{E}\{e^{kG}\}=e^{k^{2}/2}$}$ for $\mbox{$G\sim\mathcal{N}(0,1)$}$,
(\ref{eq:A-step1}) simplifies to\[
A\,\mathop\sim\limits ^{{\scriptscriptstyle \mathrm{a.s.}}}\,\mathcal{S}\left(\alpha=\frac{1}{b},\:\beta=1,\:\gamma=\lambda\pi C_{1/b}^{-1}e^{2\sigma^{2}/b^{2}}\right),\]
 for $\mbox{$b>1$}$. This is the result in (\ref{eq:stableA}) and
the derivation is complete.

\section{Derivation of the Distribution of $\mathbf{Y}$\label{sec:Deriv-stable-Y}}

To derive the distribution of $\mathbf{Y}$ given in (\ref{eq:stableY}),
we start with the following theorem.

\begin{theorem} Let $\{\tau_{i}\}_{i=1}^{\infty}$ denote the arrival
times of a one-dimensional Poisson process with rate~$\lambda$;
let $\{\mathbf{Z}_{i}\}_{i=1}^{\infty}$ be a sequence of CS i.i.d.\ complex
r.v.'s~$\mbox{$\mathbf{Z}_{i}=Z_{i,1}+jZ_{i,2}$}$, independent of
the sequence~$\{\tau_{i}\}$ and satisfying $\mbox{$\mathbb{E}\{|\mathbf{Z}_{i}|^{\alpha}\}<\infty$}$.
If $\mbox{$0<\alpha<2$}$, then\[
\sum_{i=1}^{\infty}\frac{\mathbf{Z}_{i}}{\tau_{i}^{1/\alpha}}\,\mathop\sim\limits ^{{\scriptscriptstyle \mathrm{a.s.}}}\,\mathcal{S}_{\mathrm{c}}\left(\alpha,\:\beta=0,\:\gamma=\lambda C_{\alpha}^{-1}\mathbb{E}\{|Z_{i,n}|^{\alpha}\}\right),\]
where $C_{\alpha}$ is defined in (\ref{eq:Cx}).

\end{theorem}

\begin{proof}See \cite{SamTaq:94}. For an alternative proof based
on the influence function method, see \cite{Zol:86}.\end{proof}

Using the Poisson mapping theorem as in Appendix~\ref{sec:Deriv-stable-A},
we can apply the above theorem to (\ref{eq:ysum}) and write\begin{multline}
\mathbf{Y}=\sum_{i=1}^{\infty}\frac{e^{\sigma G_{i}}\mathbf{X}_{i}}{R_{i}^{b}}=\sum_{i=1}^{\infty}\frac{\overbrace{e^{\sigma G_{i}}\mathbf{X}_{i}}^{\mathrm{CS\; i.i.d.}}}{(\underbrace{R_{i}^{2}}_{\tau_{i}})^{b/2}}\\
\,\mathop\sim\limits ^{{\scriptscriptstyle \mathrm{a.s.}}}\,\mathcal{S}_{\mathrm{c}}\left(\alpha=\frac{2}{b},\:\beta=0,\:\gamma=\lambda\pi C_{2/b}^{-1}\mathbb{E}\{|e^{\sigma G_{i}}X_{i,n}|^{2/b}\}\right),\label{eq:Y-step1}\end{multline}
for $b>1$. Note that $\mathbf{X}_{i}$, whose expression is given
in (\ref{eq:xi}), is CS due to the uniform phase~$\phi_{i}$. As
a result, $e^{\sigma G_{i}}\mathbf{X}_{i}$ is also CS. Using the
moment property of log-normal r.v.'s, i.e., $\mbox{$\mathbb{E}\{e^{kG}\}=e^{k^{2}/2}$}$
with $\mbox{$G\sim\mathcal{N}(0,1)$}$, (\ref{eq:Y-step1}) simplifies
to\[
\mathbf{Y}\,\mathop\sim\limits ^{{\scriptscriptstyle \mathrm{a.s.}}}\,\mathcal{S}_{\textrm{c}}\left(\alpha=\frac{2}{b},\:\beta=0,\:\gamma=\lambda\pi C_{2/b}^{-1}e^{2\sigma^{2}/b^{2}}\mathbb{E}\{|X_{i,n}|^{2/b}\}\right),\]
for $\mbox{$b>1$}$. This is the result in (\ref{eq:stableY}) and
the derivation is complete.

\section*{Acknowledgements}

\noindent The authors would like to thank L.~Greenstein, J.~H.~Winters,
G.~J.~Foschini, M.~Chiani, and A.~Giorgetti for their helpful
suggestions.

\bibliographystyle{../bibtex/IEEEtran}
\bibliography{../bibtex/IEEEabrv,../bibtex/StringDefinitions,../bibtex/WGroup,../bibtex/BiblioCV}

\end{document}